\documentclass[
  aps,
  prd,
  twocolumn,
  amsmath,
  amssymb,
  superscriptaddress,
  nofootinbib,
  floatfix
]{revtex4-2}

\usepackage{tikz}
\usetikzlibrary{arrows.meta}
\definecolor{gwpurple}{HTML}{7F77DD}
\definecolor{gwteal}{HTML}{1D9E75}
\definecolor{gwcoral}{HTML}{D85A30}
\definecolor{gwblue}{HTML}{378ADD}
\newcommand{\gwgeodesic}[3][]{%
  \pgfmathsetmacro{\gwA}{#2}%
  \pgfmathsetmacro{\gwS}{#3}%
  \pgfmathsetmacro{\gwR}{abs(tan(\gwS/2))}%
  \ifdim\gwS pt<180pt
    \pgfmathsetmacro{\gwStart}{\gwA+270}\pgfmathsetmacro{\gwEnd}{\gwA+90+\gwS}%
  \else
    \pgfmathsetmacro{\gwStart}{\gwA+90}\pgfmathsetmacro{\gwEnd}{\gwA+\gwS-90}%
  \fi
  \draw[#1] (\gwA:1) arc[start angle=\gwStart, end angle=\gwEnd, radius=\gwR];%
}

\usepackage{graphicx}
\usepackage{amsfonts}
\usepackage{tensor}
\usepackage{braket}
\usepackage{enumitem}
\usepackage[dvipsnames]{xcolor}
\usepackage{comment}
\usepackage{hyperref}
\usepackage[capitalise,noabbrev]{cleveref}   

\newcommand{\Sg}{S_{\mathrm{gen}}}

\newcommand{\Ar}[1]{\mathrm{Area}[#1]}

\usepackage{amsthm, mdframed}
\newmdtheoremenv[linewidth=0.6pt, innerleftmargin=8pt, innerrightmargin=8pt]{theorem}{Theorem}

\begin{document}
\title{Generalized entanglement wedges in three dimensions }

\author{Abhisek Sahu}
\email{abhi@phas.ubc.ca}
\affiliation{Department of Physics and Astronomy, University of British Columbia,
6224 Agricultural Road, Vancouver, B.C., V6T 1Z1, Canada}

\date{\today}

\begin{abstract}
Generalized entanglement wedges (GWs), proposed by Bousso and Penington for gravitating regions in general spacetimes, share many properties of ordinary entanglement wedges, including nesting, monotonicity and strong subadditivity of their generalized entropies. In AdS/CFT, the corresponding properties of entanglement wedges are consequences of the structure of the dual CFT. In this paper we ask whether the properties of GWs have a similar origin. We explore this question in vacuum AdS$_3$ at leading order in $G$. We first show that every GW on a static slice is a union of geodesically convex regions. Each such region is an intersection of ordinary entanglement wedges, so every GW is built from unions and intersections of entanglement wedges. We then use the Crofton formula of integral geometry to write the perimeter of a convex region as the measure of the set of geodesics that intersect it. In AdS$_3$ every geodesic is the Ryu--Takayanagi surface of a boundary interval, so this measure can be written in terms of the von Neumann entropies of CFT intervals. Strong subadditivity of the CFT entropy guarantees that the measure is positive. Monotonicity and strong subadditivity of GWs then follow from simple inequalities between sets of geodesics. The strong subadditivity of GWs is therefore a direct consequence of strong subadditivity on the boundary. Our results further suggest that GWs correspond to subsystems of the CFT that are more general than boundary subregions.
\end{abstract}

\maketitle

\section{Introduction}
\label{sec:BPWedge}

 A special feature of the AdS/CFT dictionary is that it holds not only between the full theories but also between their subsystems. Each subregion $A$ of the boundary is associated with a bulk region, its entanglement wedge $\mathcal{E}(A)$ \cite{Ryu:2006bv, Hubeny:2007xt, Czech:2012bh, Wall:2012uf, Headrick:2014cta}. Entanglement wedge reconstruction states that bulk operators in $\mathcal{E}(A)$ can be represented by boundary operators supported in $A$ \cite{Jafferis:2015del, Dong:2016eik}. More recently, it has been argued that in the $G \rightarrow 0$ limit the algebra of bulk operators in an entanglement wedge is dual to the algebra of boundary operators in the corresponding subregion \cite{Leutheusser:2021frk, Leutheusser:2022bgi}. The geometry of entanglement wedges mirrors the structure of these subsystems. Nested subregions have nested wedges, and spacelike separated subregions have spacelike separated wedges. The Ryu--Takayanagi formula and its quantum generalizations identify the generalized entropy of $\mathcal{E}(A)$ with the von Neumann entropy of $A$ \cite{Ryu:2006bv, Hubeny:2007xt, Faulkner:2013ana, Engelhardt:2014gca}. These ideas show that entanglement wedges encode the mathematical structure of the dual quantum theory.

For asymptotically flat spacetimes, de Sitter space and generic cosmological spacetimes, we do not yet have an example of a complete theory of quantum gravity. As a result, our understanding of the mathematical structure of quantum gravity in such spacetimes is fairly limited. We do not know what the fundamental degrees of freedom are, how they are organized into subsystems, or what structure these subsystems possess. It is also not clear which regions of such a spacetime should be associated with independent subsystems.
 
Motivated by the question of how degrees of freedom and information are organized in gravitating regions, Bousso and Penington proposed that any gravitating region has an associated region containing it, called its generalized entanglement wedge (GW) \cite{Bousso:2022hlz, Bousso:2023sya}. While GWs can be defined in any spacetime without reference to a conformal boundary, they share many properties of ordinary entanglement wedges. The GWs of nested regions are nested. Their generalized entropies are monotonic under inclusion, in the sense that a GW has generalized entropy no smaller than that of any GW it contains with the same conformal boundary \cite{Sahu:2025upe}. They also satisfy strong subadditivity  \cite{Bousso:2022hlz} for three non-overlapping regions. Moreover, for mutually independent input regions, GWs satisfy the monogamy of mutual information and the other inequalities of the static holographic entropy cone \cite{Bousso:2025mdp}.

In AdS/CFT, properties of this kind are consequences of the subsystem structure of the CFT and of its states. For example, strong subadditivity of the generalized entropy of entanglement wedges follows, via the RT formula, from strong subadditivity of the von Neumann entropy in the CFT \cite{ Headrick:2007km}. It is therefore plausible that, in a general spacetime, the properties of GWs reflect the mathematical structure of the as yet unknown fundamental theory. This possibility motivates the present paper.

In this paper we investigate the properties of GWs in asymptotically AdS spacetimes. Even in asymptotically AdS spacetimes, the family of GWs is considerably larger than the family of entanglement wedges. It includes bounded regions that do not reach the conformal boundary, as well as regions that reach the boundary but are not the entanglement wedge of any boundary subregion. In this setting, however, the fundamental description is the dual CFT. It is therefore instructive to ask whether the properties of GWs can be traced to properties of the CFT. Such a connection would give hints about the kind of structure to look for in a fundamental description of more general spacetimes.
 
We address this question in vacuum AdS$_3$, where the analysis can be made explicit. We work on a static slice, the hyperbolic disk $\mathbb{H}^2$, and at leading order in $G$, where the generalized entropy of a region is the length of its boundary in units of $4G$.  We begin in \Cref{sec:properties} by reviewing the definition of generalized entanglement wedges and showing that they satisfy axioms of an order-theoretic lattice\footnote{In order theory, a lattice is a partially ordered set in which every pair of elements has a least upper bound, called the join, and a greatest lower bound, called the meet. } . In \Cref{sec:characterization} we characterize all GWs in vacuum AdS$_3$. We show that every GW can be constructed from unions and intersections of ordinary entanglement wedges and their complements.
 We then recast this characterization in \Cref{sec:kinematic} within \textit{kinematic space}, the space of entanglement wedges of  $\mathbb{H}^2$ \cite{Czech:2014ppa, Czech:2015qta}.  Each GW is described by a corresponding \textit{wedge set} in kinematic space, and the lattice operations on GWs are viewed as lattice operations on wedge sets.
In \Cref{sec:entropy} we introduce the Crofton measure on the kinematic space. This measure can be written in terms of the von Neumann entropies of boundary intervals \cite{Czech:2015qta}.Its density is the conditional mutual information of adjacent intervals, and its positivity follows from strong subadditivity in the CFT. We show that the monotonicity and strong subadditivity of GWs follow from the positivity of the measure. They are therefore consequences of strong subadditivity of von Neumann entropy in the dual CFT. The argument does not require a density matrix or an algebra for any bounded bulk region.
Finally, \Cref{sec:discussion} discusses the assumptions and limitations of our analysis, and the interpretation of our results.

\paragraph*{Note on related work:} Several recent works have approached the quantum mechanical description of a gravitating region directly, by assigning it an algebra, by computing its entropy from the gravitational path integral, or by assigning it a quantum state. In \cite{Sahu:2025upe}, it was proposed that each GW is associated with a subalgebra of the fundamental description. Together with a proposed algebraic generalization of the RT formula, monotonicity and strong subadditivity of GWs then follow from inequalities satisfied by algebraic entropies. Kaya, Rath and Ritchie derived the Bousso--Penington proposal in time reflection symmetric settings from the gravitational path integral \cite{Kaya:2025vof}. Their derivation is motivated by random tensor networks, in which the entropy of a bulk region is defined by removing the tensors in the region and computing the entropy of the resulting open legs. Balasubramanian and Cummings computed the entropy of an arbitrary region of a gravitating spacetime using a path integral over its embeddings, interpreted as a sum over edge modes, and found agreement with the Bousso--Penington prescription \cite{Balasubramanian:2023dpj}. They have since constructed diffeomorphism invariant tensor networks for three dimensional gravity, in which subregion entropies defined through edge modes reproduce the holographic entropy formula \cite{Balasubramanian:2025rcr, Balasubramanian:2026vlr}.
In \cite{Basu:2026hbg}, the authors construct tensor networks on the geodesics of kinematic space that lie within a bulk subregion. These networks model the surface/state correspondence and generalized entanglement wedges in vacuum AdS$_3$. Their construction uses the same kinematic space and Crofton formula as Sections \ref{sec:kinematic} and \ref{sec:entropy}. It is aimed, however, at assigning a quantum state to a region, rather than at the entropic properties of GWs.
A further line of work assigns a pure quantum state directly to a gravitating region, prepared by a gravitational path integral \cite{Bousso:2026btw, Bousso:2026txw, Wei:2026syv, Anikeeva:2026zfq}. The entropies of subsystems of this state are computed by minimal surface prescriptions, which in \cite{Bousso:2026btw} agree with the Bousso--Penington prescription. Our results do not assign a state, an algebra or a path integral to a GW. They relate the properties of GWs to the structure of the underlying CFT, and neither rely on nor rule out any of the approaches above.

\section{Properties of generalized wedges}
\label{sec:properties}

Consider a Cauchy slice $\Sigma$ of a static spacetime $\mathcal{M}$. The spacetime may or may not have a conformal boundary. 
For an open subset $A \subset \Sigma$, Bousso and Penington defined the generalized entanglement wedge $E(A)$ of $A$ as the regular open subset with the same conformal boundary as $A$, containing $A$ with the smallest generalized entropy. 
Here the generalized entropy of a region is defined as the usual sum
\begin{equation}
    S_{gen}(A) = \frac{\Ar{\partial A}}{4G} + S_{QFT}(A),
\end{equation}
of the boundary area\footnote{Note that here, $\partial A$ denotes only the part of the boundary of $A$ in the interior of the slice $\Sigma$.} of $A$ and the entropy of any quantum field theory degrees of freedom residing in $A$. 

In this paper we will find it useful to refer to properties of generalized wedges without any reference to seed regions. 
An equivalent definition of GWs is:

\medskip
\noindent\textbf{Definition:}\textit{ A \textbf{generalized entanglement wedge} $W$ is a regular open subset
of $\Sigma$ with the
property that no region containing W and having the same conformal boundary as W has a smaller generalized
entropy.}

The definition of GWs above implies the following geometric properties  which promote the set of generalized wedges to a lattice. 
\begin{itemize}
    \item \textbf{Partial order}: The set of GWs is partially ordered by inclusion.
    \item \textbf{Meet}: For any two wedges $W_1, W_2$, the geometric intersection $W_1 \cap W_2$ is also a wedge. So the \textit{meet} operation is same as intersection.
    \item  \textbf{Join}: 
    The \textit{join} of two GWs $W_1, W_2$ is defined as the smallest wedge that contains $W_1 \cup W_2$. Note that the geometric union of $W_1, W_2$ is not a GW generally, and hence the join and the union of two wedges generically differ. 
\end{itemize}
 Additionally, the generalized entropy of GWs satisfy the following properties:
 \begin{itemize}
     \item \textbf{Monotonicity}: For GWs $W_1 \subseteq W_2, \;\Sg(W_1) \leq \Sg(W_2)$.
     \item \textbf{Strong Subadditivity}: For GWs  $W_1, W_2$, \begin{align*}
         \Sg(W_1)
    &+ \Sg(W_2) 
    \geq \\
    &\Sg(W_1 \vee W_2)
    + \Sg(W_1 \land W_2).
     \end{align*}
 \end{itemize}

 While the above properties directly follow from the definition of GWs in bulk spacetime, the goal of this paper is to ask whether there is an explanation from the perspective of the boundary field theory.

\section{Characterization of generalized entanglement wedges}
\label{sec:characterization}

\begin{figure*}
  \centering
\begin{tikzpicture}[every node/.style={font=\small}]
  \begin{scope}[scale=2.1]
    \draw[thick, black!70] (0,0) circle (1);
    \fill[gwteal, fill opacity=0.25]
      plot[domain=0:360, samples=240, smooth cycle] (\x:{0.5*(1+0.55*cos(2*\x))});
    \fill[gwcoral, fill opacity=0.4] plot coordinates {(-0.224,0.334) (-0.113,0.321) (0.000,0.317) (0.113,0.321) (0.224,0.334) (0.221,0.329) (0.178,0.313) (0.133,0.291) (0.036,0.235) (0.001,0.225) (-0.015,0.227) (-0.033,0.234) (-0.143,0.296) (-0.221,0.329)} -- cycle;
    \fill[gwcoral, fill opacity=0.4] plot coordinates {(-0.224,-0.334) (-0.113,-0.321) (-0.000,-0.317) (0.113,-0.321) (0.224,-0.334) (0.221,-0.329) (0.178,-0.313) (0.133,-0.291) (0.036,-0.235) (0.001,-0.225) (-0.015,-0.227) (-0.033,-0.234) (-0.143,-0.296) (-0.221,-0.329)} -- cycle;
    \draw[gwteal, thick]
      plot[domain=0:360, samples=240, smooth cycle] (\x:{0.5*(1+0.55*cos(2*\x))});
    \draw[thick] plot coordinates {(-0.224,0.334) (-0.113,0.321) (0.000,0.317) (0.113,0.321) (0.224,0.334)};
    \draw[thick] plot coordinates {(-0.224,-0.334) (-0.113,-0.321) (-0.000,-0.317) (0.113,-0.321) (0.224,-0.334)};
    \node at (-0.42,0) {$A$};
    \node[anchor=south] at (0,0.344) {$S$};
    \node[anchor=north] at (0,0.215) {$S'$};
    \node at (0,-1.22) {(a)};
  \end{scope}
  \begin{scope}[shift={(5,0)}, scale=2.1]
    \draw[thick, black!70] (0,0) circle (1);
    \draw[black!80, dashed] plot coordinates {(-0.830,0.000) (-0.824,-0.055) (-0.808,-0.109) (-0.782,-0.157) (-0.748,-0.198) (-0.705,-0.233) (-0.656,-0.259) (-0.605,-0.274) (-0.550,-0.280) (-0.506,-0.277) (-0.375,-0.250) (-0.247,-0.232) (-0.129,-0.222) (-0.011,-0.218) (0.118,-0.221) (0.247,-0.232) (0.359,-0.248) (0.524,-0.279) (0.565,-0.280) (0.605,-0.274) (0.650,-0.261) (0.693,-0.241) (0.733,-0.212) (0.766,-0.178) (0.792,-0.140) (0.812,-0.098) (0.826,-0.050) (0.830,-0.003) (0.825,0.052) (0.809,0.106) (0.783,0.155) (0.750,0.196) (0.707,0.232) (0.659,0.258) (0.605,0.274) (0.550,0.280) (0.506,0.277) (0.375,0.250) (0.247,0.232) (0.129,0.222) (0.011,0.218) (-0.118,0.221) (-0.247,0.232) (-0.359,0.248) (-0.524,0.279) (-0.565,0.280) (-0.605,0.274) (-0.653,0.260) (-0.695,0.240) (-0.733,0.212) (-0.766,0.178) (-0.792,0.140) (-0.813,0.095) (-0.826,0.050) (-0.830,0.003)} -- cycle;
    \fill[gwteal, fill opacity=0.25] (-0.55,0) circle (0.28) (0.55,0) circle (0.28);
    \draw[gwteal, thick] (-0.55,0) circle (0.28) (0.55,0) circle (0.28);
    \node at (-0.55,0) {$K_1$};
    \node at (0.55,0) {$K_2$};
    \node at (0,-1.22) {(b)};
  \end{scope}
  \begin{scope}[shift={(10,0)}, scale=2.1]
    \draw[thick, black!70] (0,0) circle (1);
    \draw[black!80, dashed] plot coordinates {(-0.580,0.000) (-0.575,-0.053) (-0.560,-0.102) (-0.535,-0.149) (-0.501,-0.191) (-0.459,-0.225) (-0.412,-0.250) (-0.363,-0.265) (-0.310,-0.270) (-0.147,-0.253) (-0.027,-0.248) (0.125,-0.252) (0.304,-0.270) (0.347,-0.268) (0.388,-0.258) (0.428,-0.243) (0.464,-0.222) (0.497,-0.195) (0.525,-0.163) (0.548,-0.128) (0.566,-0.086) (0.576,-0.045) (0.580,-0.003) (0.576,0.048) (0.561,0.099) (0.536,0.147) (0.503,0.189) (0.462,0.223) (0.415,0.249) (0.363,0.265) (0.310,0.270) (0.147,0.253) (0.027,0.248) (-0.125,0.252) (-0.304,0.270) (-0.347,0.268) (-0.388,0.258) (-0.428,0.243) (-0.464,0.222) (-0.497,0.195) (-0.525,0.163) (-0.548,0.128) (-0.565,0.089) (-0.576,0.048) (-0.580,0.003)} -- cycle;
    \fill[gwteal, fill opacity=0.25] (-0.31,0) circle (0.27) (0.31,0) circle (0.27);
    \draw[gwteal, thick] (-0.31,0) circle (0.27) (0.31,0) circle (0.27);
    \node at (-0.31,0) {$K_1$};
    \node at (0.31,0) {$K_2$};
    \node[anchor=south] at (0,0.290) {$K_1 \vee K_2$};
    \node at (0,-1.22) {(c)};
  \end{scope}
\end{tikzpicture}
  \caption{Examples for the characterization theorem. Shaded regions are the
candidate wedges, and dashed curves are convex hulls. (a) A nonconvex region
$A$. Each boundary arc $S'$ can be replaced by the geodesic segment $S$ with
the same endpoints. Adding the region between them yields a larger region with
smaller perimeter, so $A$ is not a generalized entanglement wedge.
(b) Two convex regions far apart. The hull of $K_1 \cup K_2$ has a larger
perimeter than $K_1$ and $K_2$ together, so $K_1 \cup K_2$ is a generalized
entanglement wedge and $K_1 \vee K_2 = K_1 \cup K_2$. (c) Two convex regions
close together. The hull has a smaller perimeter than $K_1$ and $K_2$
together, so $K_1 \cup K_2$ is not a generalized entanglement wedge. Its
generalized entanglement wedge is the hull, $K_1 \vee K_2$.}
  \label{fig:char}
\end{figure*}
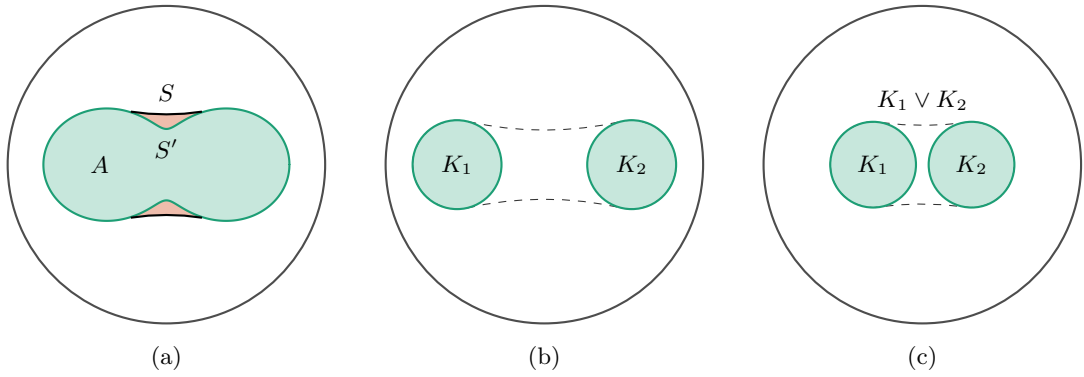

In this section, we consider a time slice of vacuum anti-de Sitter spacetime (a hyperbolic disk) and describe the various spatial subsets that qualify as GWs. 

A generalized wedge in $H^2$ is an open subset $W$  that is the interior of its closure, such that there is no region $R$ with $W \subset R$, $\tilde{\partial} W = \tilde{\partial} R$, and $S_{\rm gen}(R) - S_{\rm gen}(W) < 0$. Here $\tilde{\partial}$ denotes the conformal boundary. We restrict to wedges whose closure lies in the interior of $H^2$, so that $\tilde{\partial} W = \emptyset$ and all perimeters are finite. Since we are considering vacuum AdS, we expect that the area term in the generalized entropy dominates, so the last condition can be replaced by ${\rm Per}(R) - {\rm Per}(W) < 0$. When more than one region containing $W$ has the least perimeter, we take $E(W)$ to be the smallest of these under inclusion, following~\cite{Bousso:2022hlz}.
We now show the following general result \footnote{The proof below was first written down by Mark Van Raamsdonk.}:

\vspace{2mm}
\noindent\fbox{\parbox{\dimexpr\linewidth-2\fboxsep-2\fboxrule\relax}{
{\bf Theorem:} {\it A wedge in $H^2$ is a generalized entanglement wedge if and only if each of its connected components is geodesically convex, and the convex hull of any collection of these components has a perimeter larger than or equal to that of the collection.}}}
\vspace{2mm}

We will refer to this result as the \textit{characterization theorem} from here on. For an example of possible GWs, see \cref{fig:char}.

Let $W$ be a generalized entanglement wedge. We first show that every connected component $A$ of $W$ must be geodesically convex. If $A$ were not geodesically convex, 
there would be a geodesic segment $S$ connecting two points on the boundary of $A$ such that $S$ intersects $A$ only at these two points. The portion $S'$ of the boundary of $A$ connecting these two points is not $S$, and therefore not geodesic (there is a unique geodesic segment connecting any two points in $H^2$). By adding the region bounded by $S$ and $S'$ to $A$ and thus to $W$, we obtain a set that contains $W$ but has smaller boundary length. This remains true if the added region meets other components of $W$, since the new boundary is contained in the old boundary with $S'$ replaced by $S$. This is not possible for a GW. Next, suppose that $W$ has a collection of components whose convex hull has smaller perimeter than these components. By replacing the components with their convex hull, we would then form a set containing $W$ with a smaller perimeter, even if the hull meets other components of $W$. This again contradicts the assumption that $W$ is a GW.

To prove the sufficiency part, it will be useful to note the following basic geometrical fact:

\vspace{2mm}
\noindent
{\bf Lemma} {\it If $A \subset B$ are both geodesically convex subsets of $H^2$, then the perimeters satisfy ${\rm Per}(B) - {\rm Per}(A) \ge 0$.}
\vspace{2mm}

One way to see this is using the Crofton formula: there exists a measure $\mu$ on the space ${\cal L}$ of all complete geodesics in $H^2$ such that the perimeter of any convex bounded region $K$ is 
\[
{\rm Per}(K) = \mu(\{\ell \in {\cal L} | \ell \cap K \ne \emptyset \}) \; ,
\]
i.e. the measure of the set of geodesics that intersect $K$.
If $K_1 \subset K_2$, the set on the right grows and therefore has larger measure. This implies ${\rm Per}(K_2) - {\rm Per}(K_1) \ge 0$. In \cref{sec:entropy} we will write an explicit formula for the Crofton form using AdS/CFT.

Now, suppose that each connected component of $W$ is geodesically convex, that the convex hull of any collection of these components has a perimeter larger than or equal to that of the collection, and suppose further that $W$ is {\it not} a GW.
In this case, $E(W)$ is a GW that contains $W$ as a proper subset. 
By the above, the connected components of $E(W)$ are geodesically convex. Each connected component of $W$ lies in a single component of $E(W)$. A component of $E(W)$ containing no point of $W$ could be removed to lower the perimeter, so every component of $E(W)$ meets $W$. Since $E(W)$ is strictly larger than $W$, one of these connected components (call it $K_1$) contains as a proper subset a set $K_0 = W \cap K_1$ that is a union of the connected components of $W$. Replacing $K_1$ with $K_0$ therefore gives a set that still contains $W$.
The perimeter of $K_0$ must be strictly larger than the perimeter of $K_1$, otherwise we could replace $K_1$ with $K_0$ to get a smaller set containing $W$ with smaller or equal generalized entropy, contradicting the definition of $E(W)$.
Since $K_1$ is geodesically convex, it contains the convex hull of $K_0$. According to the lemma, the perimeter of $K_1$ is larger than or equal to the perimeter of the convex hull of $K_0$. According to our assumptions, the convex hull of $K_0$ has a perimeter that is larger than or equal to the perimeter of $K_0$. Thus, the perimeter of $K_1$ is larger than or equal to the perimeter of $K_0$. This contradicts the previous paragraph, and hence $W$ is a GW.

A couple of further results allow us to characterize GWs in terms of standard entanglement wedges. On a time slice of vacuum AdS$_3$, the entanglement wedge of a boundary interval is the open region between the interval and its RT geodesic, i.e. an open hyperbolic half-plane.

\vspace{1mm}
\noindent
\textbf{Proposition 1:}
{\it A geodesically convex connected open set in $H^2$ is equal to the intersection of all the ordinary entanglement wedges that contain it.}
\vspace{1mm}

\noindent
\textbf{Proposition 2:}
{
\it A geodesically convex connected open set in $H^2$ is equal to the interior of the intersection of the complement of all the entanglement wedges that do not intersect it.
}
\vspace{1mm}

Both follow from the separation property of convex sets in $H^2$. For any point $x$ outside a convex open set $K$, there is a geodesic separating $x$ from $K$, or a supporting geodesic of $K$ through $x$ if $x$ lies on the boundary of $K$. The open half-plane on the side of $K$ is an entanglement wedge that contains $K$ but not $x$. This gives Proposition 1. The complements in Proposition 2 are closed half-planes, so their intersection is the closure of $K$. Its interior is $K$ itself. The two statements are related by complementarity, since the complement of the entanglement wedge of an interval is the closure of the entanglement wedge of the complementary interval.

\section{Kinematic space}
\label{sec:kinematic}

\begin{figure*}
  \centering
\begin{tikzpicture}[
    >={Stealth[length=2.2mm]},
    every node/.style={font=\small},
  ]
  \def\pz{210}      
  \def\rK{0.41421}  
  \begin{scope}[scale=2.9]
    \fill[gwpurple, fill opacity=0.12]
      (\pz:1) arc[start angle=\pz+90, end angle=\pz+270-90, radius=1]
              arc[start angle=\pz+270, end angle=\pz, radius=1] -- cycle;
    \draw[thick, black!70] (0,0) circle (1);
    \fill[gwteal, fill opacity=0.25] (0,0) circle (\rK);
    \draw[gwteal, thick] (0,0) circle (\rK);
    \node at (-0.13,0.2) {$K$};
    \gwgeodesic[->, thick]{\pz}{270}
    \gwgeodesic[->, thick]{\pz}{90}
    \gwgeodesic[->, gwpurple, thick, dashed]{\pz}{315}
    \gwgeodesic[->, gwteal,   thick, dashed]{\pz}{200}
    \gwgeodesic[->, gwcoral,  thick, dashed]{\pz}{45}
    \draw[->, black!55] (\pz:1.1) arc[start angle=\pz, end angle=\pz+270, radius=1.1];
    \node at (\pz+135:1.2) {$s_K$};
    \fill (\pz:1) circle (0.022);
    \node[anchor=east] at (\pz:1.03) {$p_0$};
    \node[anchor=south east] at (\pz+270:1.13) {$p_0+s_K$};
    \node[anchor=north west] at (\pz+90:1.13) {$p_0+2\pi-s_K$};
  \end{scope}

  \begin{scope}[shift={(5.6,-3)}]
    \def\L{6}
    \fill[gwpurple, fill opacity=0.22] (0,4.5) rectangle (\L,6);
    \fill[gwteal,   fill opacity=0.22] (0,1.5) rectangle (\L,4.5);
    \fill[gwcoral,  fill opacity=0.22] (0,0)   rectangle (\L,1.5);
    \draw[gwpurple, thick] (0,4.5) -- (\L,4.5);
    \draw[gwcoral,  thick] (0,1.5) -- (\L,1.5);
    \draw[black!40, densely dotted] (0,3) -- (\L,3);
    \draw[black!60, dashed] (0,0) -- (\L,0);
    \draw[black!60, dashed] (0,6) -- (\L,6);
    \draw[black!60] (0,0) -- (0,6);
    \draw[black!60, dotted] (\L,0) -- (\L,6);
    \pgfmathsetmacro{\xz}{\L*\pz/360}
    \draw[black!70] (\xz,0) -- (\xz,6);
    \node[anchor=south] at (\xz,6) {$p_0$};
    \fill[gwpurple] (\xz,{\L*315/360}) circle (0.07);
    \fill[gwteal]   (\xz,{\L*200/360}) circle (0.07);
    \fill[gwcoral]  (\xz,{\L*45/360})  circle (0.07);
    \node at (1.4,5.25) {$W(K)$};
    \node at (1.4,3.6)  {$\Gamma_K$};
    \node at (1.4,0.75) {$W^{T}(K)$};
    \foreach \y/\lab in {0/0, 1.5/2\pi-s_K, 3/\pi, 4.5/s_K, 6/2\pi}
      \node[anchor=east] at (-0.08,\y) {$\lab$};
    \foreach \x/\lab in {0/0, 3/\pi, 6/2\pi}
      \node[anchor=north] at (\x,-0.08) {$\lab$};
    \node at (3,-0.75) {$p$};
    \node[anchor=south] at (0,6.15) {$s$};
  \end{scope}
\end{tikzpicture}
\caption{A convex region and its kinematic space. Left: the region $K$ is a circular region centred in the hyperbolic disk. From a boundary point $p_0$,
the two solid black arrows represent oriented geodesics tangent to $K$. 
The entanglement wedge corresponding to a geodesic lies on its anti-clockwise side.
The upper black geodesic bounds the grey shaded
wedge $w(p_0, p_0+s_K)$, the smallest wedge from $p_0$ that contains $K$.
The dashed geodesics are samples.  The green geodesic cuts $K$, the entanglement wedge bounded by the purple geodesic includes $K$, where as the same for the orange geodesic does not contain K.
Right: Kinematic space in the coordinates $(p, s)$, with $s = q - p$. Since
$K$ is centred, the critical angle $s_K$ is independent of
$p$, and kinematic space splits into three strips. These are the wedge set
$W(K)$, the intersect set $\Gamma_K$ and the transposed wedge set $\tilde{W}(K)$.
The dots on the line $p = p_0$ mark the three sample geodesics in matching colours. The edges $s = 0$ and $s = 2\pi$ are excluded, and the edges
$p = 0$ and $p = 2\pi$ are identified. The transposition
$(p, q) \to (q, p)$ acts as a reflection about the dotted line $s = \pi$ and
exchanges $W(K)$ with $\tilde{W}(K)$. }
\label{fig:kinematic}
\end{figure*}
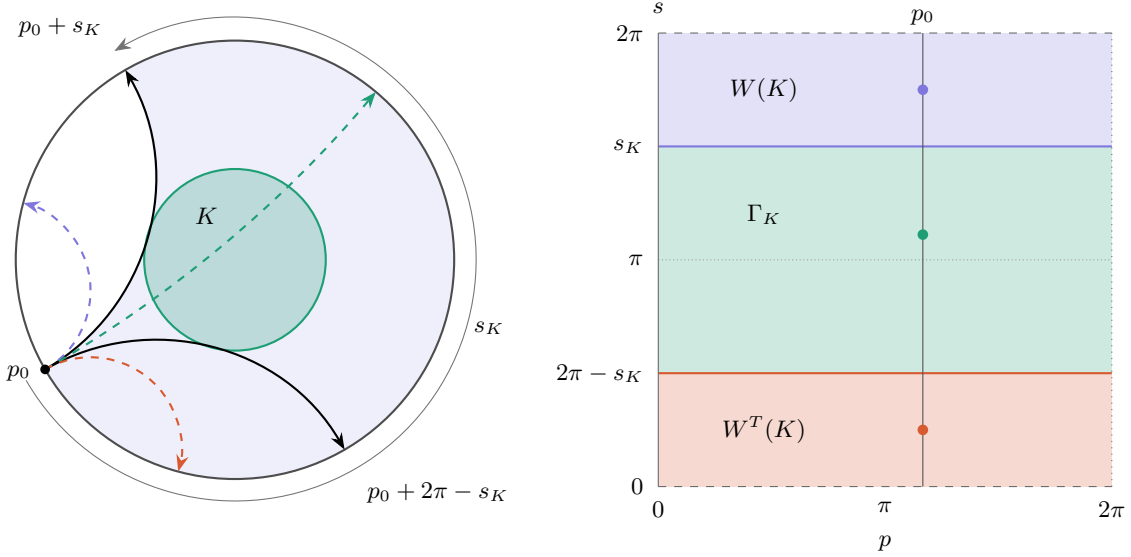

The characterization of Sec.~\ref{sec:characterization} describes a
generalized wedge as a union of geodesically convex sets, each of which
is an intersection of ordinary entanglement wedges. It is therefore
natural to trade the description of a GW in terms of the bulk points it
contains for a description in terms of the ordinary entanglement wedges
that contain it. In this section we set up a description of the space of entanglement wedges, called a kinematic space. We then show that the lattice structure of Sec.~\ref{sec:properties} is faithfully represented in the kinematic space. 
The
generalized entropy can then be written as an integral over the kinematic space, which is later used in  Sec.~\ref{sec:entropy} for a boundary interpretation of the entropic properties of GWs.

\subsection{Kinematic space: the space of complete geodesics}

A complete geodesic of $H^2$ is fixed by its two endpoints on the
conformal boundary. Labelling a geodesic by the ordered pair of boundary
angles $\ell=(p,q)$, the space of geodesics
\begin{equation}
  \Gamma=\{(p,q)\in S^1\times S^1 : p\neq q\}
\end{equation}
is two dimensional and is known as the kinematic space of the slice. We write $w(\ell)$ for the entanglement
wedge of the interval $(p,q)$, namely the half-plane bounded by $\ell$
and by the boundary arc running counterclockwise from $p$ to $q$. The
two labels $(p,q)$ and $(q,p)$ describe the same geodesic but
complementary intervals, so $w(q,p)$ is the complement of $w(p,q)$.

On a static slice of AdS$_3$ geodesics are Ryu--Takayanagi surfaces, so a
point of $\Gamma$ simultaneously represents a bulk geodesic, a boundary interval, and
an ordinary entanglement wedge.  This
triple role of $\Gamma$ is what will allow us to rewrite statements about the bulk
as statements about boundary subregions. 

The kinematic space 
$\Gamma$ also inherits a metric from the hyperbolic disk. The form of the metric is constrained by the isometries of $H^2$, which in turn also act on the space of geodesics of $H^2$, and hence the kinematic space.
Up to a normalization, this metric is the two-dimensional de Sitter metric~\cite{Czech:2015qta}.
Since $\Gamma$ is an orientable manifold,  we will use this metric to define a measure $\nu$ on the kinematic space.
Fixing the normalization using the gravitational constant, we take
\begin{equation}
  \nu(U)=\int_{U} h(p,q)\,dp\,dq,
  \;\;
  h(p,q)=\frac{1}{8G}\,\csc^{2}\!\left(\frac{q-p}{2}\right),
  \label{eq:crofton}
\end{equation}
integrated over the full range of both angles. $\nu$ is called the \textit{Crofton
measure}. 
In \cref{sec:entropy}, we will see a CFT interpretation of the Crofton form.

\subsection{ Crofton formula}

We define two subsets of $\Gamma$ related to a bulk region $K$. The first
is the intersect set: the set of geodesics that cut it,
\begin{equation}
  \Gamma_K=\{\ell\in\Gamma : \ell\cap K\neq\emptyset\}.
\end{equation}
The second is the set of ordinary entanglement wedges that contain
it,
\begin{equation}
  \mathcal{W}(K)=\{\ell\in\Gamma : K\subseteq w(\ell)\},
  \label{eq:wedgeset}
\end{equation}
which we call the wedge set of $K$. 
For a connected region these two
sets are complementary in the following sense. Any interval either has a
wedge containing $K$, or a wedge disjoint from $K$, or its geodesic cuts
$K$, and these three possibilities are exclusive. Since a wedge is
disjoint from $K$ exactly when the complementary wedge contains it,
\begin{align}
\nonumber
  \Gamma_K &= \Gamma\setminus\big(\mathcal{W}(K)\cup\tilde{\mathcal{W}}(K)\big),\\
  \tilde{\mathcal{W}}(K)&=\{(q,p):(p,q)\in\mathcal{W}(K)\}.
  \label{eq:trichotomy}
\end{align}

An example of a wedge set and an intersect set is shown in \cref{fig:kinematic}. We now describe the set $\mathcal{W}(K)$ explicitly. Fixing a boundary point
$p$, we parametrize the second endpoint by the opening angle
$s=q-p\in(0,2\pi)$. As $s$ increases, the wedge $w(p,p+s)$ grows
monotonically. It shrinks to zero for $s \rightarrow 0$ and fills the whole slice as
$s\to 2\pi$. The region $K$ is therefore contained in $w(p,p+s)$ for every
$s$ above a single critical angle,
\begin{equation}
  s_K(p)=\inf\bigl\{\,s\in(0,2\pi) : K\subset w(p,p+s)\,\bigr\}.
  \label{eq:critical-angle}
\end{equation}
It follows that
\begin{equation}
  \mathcal{W}(K)=\bigl\{(p,q) : s_K(p)\leq q-p<2\pi\bigr\}.
  \label{eq:fibre}
\end{equation}
The geodesic $\ell_{p,s_K(p)}$ is the tangent to $K$ from $p$ that
leaves $K$ on its left. The curve $p\mapsto s_K(p)$ is the boundary of
the wedge set.

A central result in integral geometry~\cite{Czech:2015qta} is the Crofton formula which computes the length of a curve through an integral in the Kinematic space. For any rectifiable curve $\gamma$,
\begin{equation}
  \frac{\mathrm{length}(\gamma)}{4G}
  =\frac{1}{4}\int_{\Gamma} d\nu\; n_\gamma ,
  \label{eq:croftonformula}
\end{equation}
where $n_\gamma(\ell)$ counts the intersections of $\gamma$ with $\ell$.
The factor of $1/4$ is fixed by the normalization of $\nu$.
A geodesic meeting a convex region crosses its boundary exactly twice,
so $n_{\partial K}=2$ on $\Gamma_K$ and vanishes elsewhere. 
In vacuum AdS, the generalized entropy of a bulk region comprises only of the length of its boundary. Thus the Crofton formula \eqref{eq:croftonformula} applied to the boundary of $K$ computes
\begin{equation}
  S_{\mathrm{gen}}(K)\simeq\frac{\mathrm{Per}(K)}{4G}
  =\tfrac{1}{2}\,\nu(\Gamma_K).
  \label{eq:SgenConvex}
\end{equation} 
As a check, a disk of hyperbolic radius $R$ about the centre of the
slice is cut by the intervals with $2\pi-2\alpha_R > q-p\geq 2\alpha_R$ and
$\cos\alpha_R=\tanh R$, and \eqref{eq:crofton} returns
$2\pi\sinh R/4G$, its generalized entropy.

A generalized wedge $W=\bigsqcup_i A_i$ with several convex components
can be described by the collection $\{\Gamma_{A_i}\}$. Since a  single geodesic
may cut more than one component, the multiplicity function is written as
\begin{equation}
  n_W=2\sum_i \mathbf{1}_{\Gamma_{A_i}}
  \label{eq:multiplicity},
\end{equation} where $\mathbf{1}_{\Gamma_{A_i}}$ takes the value $1$ for any point in $\Gamma_{A_i}$ and zero elsewhere.
Then the entropy \eqref{eq:SgenConvex} generalizes
to
\begin{equation}
  S_{\mathrm{gen}}(W)=\tfrac{1}{4}\int d\nu\; n_W
  =\sum_i \tfrac{1}{2}\,\nu(\Gamma_{A_i}).
  \label{eq:SgenGeneral}
\end{equation}

\subsection{The lattice}
\label{sec:wedge_lattice}
The lattice operations of Sec.~\ref{sec:properties} are now set
operations on wedge sets. Let $K_1,K_2$ be connected convex wedges.
\begin{itemize}
\item \emph{Partial order.} $K_1\subseteq K_2$ if and only if
  $\mathcal{W}(K_2)\subseteq\mathcal{W}(K_1)$.  The forward direction is immediate from the definition \cref{eq:wedgeset}. The reverse implication uses the result
  of Sec.~\ref{sec:characterization} that a convex region is the
  intersection of the entanglement wedges containing it, 
  \[
  K = \bigcap_{\ell \in \mathcal{W}(K)} w(\ell). \]
  When the wedge set becomes larger, the intersection contains a larger number of entanglement wedges, and hence the size of the intersection becomes smaller.
\item \emph{Join.} $\mathcal{W}(K_1\vee K_2)
  =\mathcal{W}
  (K_1)\cap\mathcal{W}(K_2)$.
  Since $K_1, K_2 \subseteq K_1 \vee K_2 $, the partial order on wedge sets tells us $\mathcal{W}(K_1\vee K_2)
  \subseteq \mathcal{W}
  (K_1)\cap\mathcal{W}(K_2)$.
  To show the reverse inclusion, we use the convexity of the entanglement wedges $w(\ell)$. Let a geodesic $\ell$ be such that $w(\ell) \supseteq K_1, K_2$ and hence $w(\ell) \supseteq K_1 \cup K_2$. Due to convexity of $w(\ell)$ it must also  contain the convex hull of $K_1\cup K_2$. Hence $\mathcal{W}(K_1 \vee K_2) \supseteq \mathcal{W}(K_1) \cap \mathcal{W}(K_2)$, completing the proof.
\item \emph{Meet.} $\mathcal{W}(K_1\wedge K_2)\supseteq
  \mathcal{W}(K_1)\cup\mathcal{W}(K_2)$. The inclusion follows again from partial ordering. The  inclusion is strict in
  general as an entanglement wedge may contain the overlap of two regions
  without containing either.
\end{itemize}
The join is the exact operation on wedge sets while the meet is not, and
in the bulk it is the other way round. This is a consequence of the
correspondence reversing inclusion. For wedges with several components
the same statements hold componentwise.

\section{The Crofton form and generalized entropy}
\label{sec:entropy}

So far the Crofton measure has been an abstract bulk entity. We now
give it a boundary interpretation and show that the entropic properties of
generalized wedges follow from strong subadditivity of the vacuum
entropy of the dual CFT.

\subsection{The Crofton measure as conditional mutual information}

We begin by recalling how kinematic space is encoded in boundary data.
On a static slice of AdS$_3$, the Ryu--Takayanagi surface of a boundary
interval is the bulk geodesic that shares its endpoints
\cite{Ryu:2006bv,Hubeny:2007xt}. The entanglement entropy of the
interval with endpoints $p$ and $q$ is therefore
\begin{equation}
  S(p,q)=\frac{\operatorname{length}(\gamma_{pq})}{4G},
  \label{eq:RT}
\end{equation}
to leading order in $G$. We set the AdS radius to one and take the
boundary circle to have unit radius. In the vacuum, conformal symmetry
fixes the entropy of an interval to be
\begin{equation}
  S(p,q)=\frac{c}{3}\log\left(\frac{2}{\epsilon}\sin\frac{q-p}{2}\right),
  \qquad c=\frac{3}{2G},
  \label{eq:Svac}
\end{equation}
where $\epsilon$ is a UV cutoff and $c$ is the Brown--Henneaux central
charge. The divergence enters only as an additive constant. It does not
depend on the positions of the endpoints.

Equation \eqref{eq:RT} identifies each point of kinematic space with a
boundary interval, and hence with an ordinary entanglement wedge. The
two orientations of a geodesic correspond to an interval and its
complement. In a pure state these have equal entropy, so
$S(p,q)=S(q,p)$. This is the boundary counterpart of the statement that
both orientations of a geodesic contribute equally to the Crofton
formula \eqref{eq:croftonformula}.

The Crofton form first appeared in holography through the differential
entropy of Balasubramanian et al.~\cite{Balasubramanian:2013lsa}. They
considered a closed convex bulk curve $\gamma$ and the one parameter
family of geodesics tangent to it. Each tangent geodesic ends on a
boundary interval $I_k=(p_k,q_k)$. The intervals cover the circle, and
consecutive intervals overlap. The differential entropy is defined as
\begin{equation}
  S_{\rm diff}=\sum_k\Big[S(I_k)-S(I_{k-1}\cap I_k)\Big].
  \label{eq:Sdiff-discrete}
\end{equation}
Each term is finite because the cutoff dependence cancels between the
two entropies. In the continuum limit the intersection
$I_{k-1}\cap I_k$ is $(p_k,q_{k-1})$, and the sum becomes
\begin{equation}
  S_{\rm diff}=\oint_{\gamma^*}\kappa,
  \qquad
  \kappa=\frac{\partial S(p,q)}{\partial q}\,dq,
  \label{eq:Sdiff}
\end{equation}
where $\gamma^*$ is the curve $q=q_\gamma(p)$ in kinematic space traced
by the tangent geodesics. Balasubramanian et al.\ showed that
$S_{\rm diff}=\operatorname{length}(\gamma)/4G$. Later work extended
this to general curves and to higher dimensions
\cite{Myers:2014jia,Headrick:2014eia,Czech:2014tva}. The one form
$\kappa$ is the primitive of the Crofton form mentioned below. Its exterior derivative is
\begin{equation}
  d\kappa=\frac{\partial^2 S(p,q)}{\partial p\,\partial q}\,dp\wedge dq .
\end{equation}
Stokes' theorem then converts \eqref{eq:Sdiff} into an integral of
$d\kappa$ over the region of kinematic space bounded by $\gamma^*$ and
the line $q=p+\pi$. In the vacuum, $\kappa$ vanishes on this line,
since $\partial_q S\propto\cot\frac{q-p}{2}$. The bounded region is
precisely the set of geodesics that cross $\gamma$, i.e $\Gamma_\gamma$. Thus the differential entropy defined above is equivalent to the
Crofton formula \eqref{eq:croftonformula}.

In~\cite{Czech:2014ppa,Czech:2015qta}, the authors made this observation the starting
point of an integral geometric reading of holography. They noted that
the Crofton density \eqref{eq:crofton} is a second derivative of the
boundary entanglement entropy,
\begin{equation}
  h(p,q)=\frac{\partial^2 S(p,q)}{\partial p\,\partial q}.
  \label{eq:hfromS}
\end{equation}
For the vacuum \eqref{eq:Svac}, a direct computation gives
\begin{equation}
  h(p,q)=\frac{c}{12}\csc^2\frac{q-p}{2}
        =\frac{1}{8G}\csc^2\frac{q-p}{2},
  \label{eq:hvac}
\end{equation}
which reproduces \eqref{eq:crofton}. The same density defines a
Lorentzian metric $ds^2=h\,dp\,dq$ on kinematic space. For the vacuum
this is two dimensional de Sitter space \cite{Czech:2015qta}. We do not
need the metric structure here. Only the volume form will play a role.

The positivity of the measure has a remarkable interpretation from the boundary perspective, by  virtue of \cref{eq:hfromS}. 
For three adjacent intervals $A=(p-dp,p)$, $B=(p,q)$ and $C=(q,q+dq)$,
\begin{align}
\nonumber
  I(A,C|B) &= S(AB)+S(BC)-S(B)-S(ABC) \\
  &=\frac{\partial^2 S(p,q)}{\partial p\,\partial q}\,dp\,dq = h(p,q)\,dp\,dq,
  \label{eq:cmi}
\end{align}
so each element of the Crofton measure is the conditional mutual
information of neighbouring boundary intervals. Strong subadditivity of
the boundary entropy is the statement $I(A,C|B)\geq 0$, and it is
therefore equivalent to positivity of $\nu$. We further note that $I(A,C|B)$ is ultraviolet finite, so $\nu$ does not
inherit the UV divergences of the von Neumann entropies from which it is
built.

We close by stating what the arguments below will use. They rely on
the identification \eqref{eq:hfromS} and the positivity of $h(p,q)$. The explicit vacuum form \eqref{eq:hvac} enters
only to identify integrals of $h$ with bulk lengths through the Crofton
formula. The bulk entropic inequalities we derive are therefore
consequences of boundary strong subadditivity, supplemented by the
Ryu--Takayanagi formula. Throughout, \eqref{eq:RT} holds at leading
order in $G$ and bulk entanglement enters at subleading order, which is
consistent with the semiclassical limit in which the generalized
entropy is dominated by the area term.

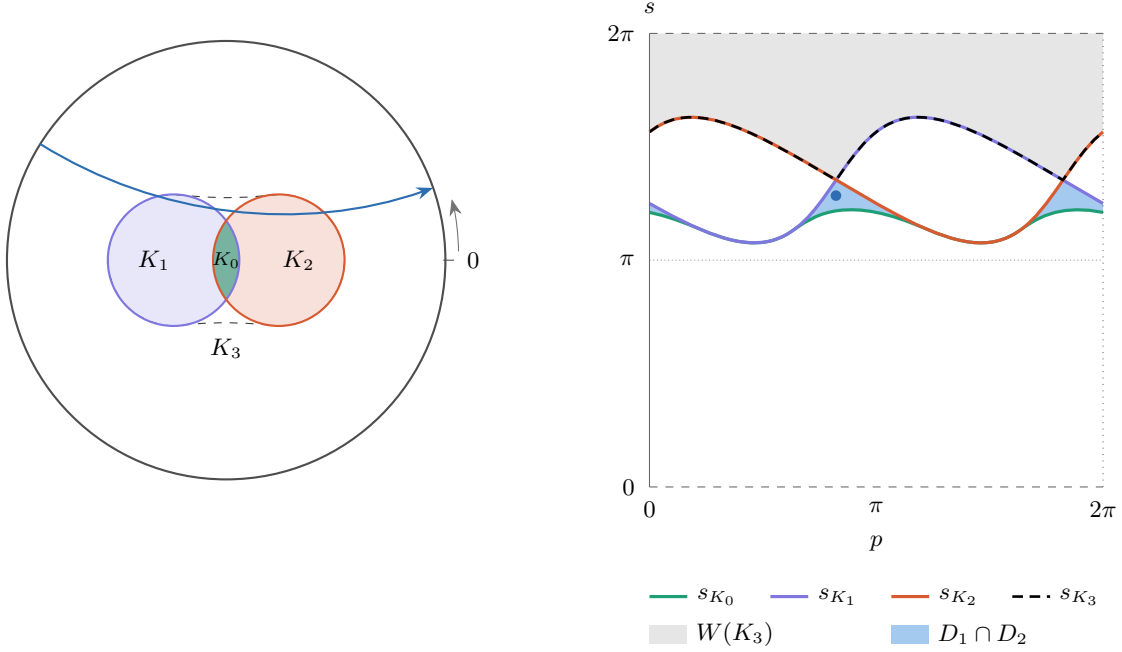
\begin{figure*}
  \centering
\begin{tikzpicture}[
    >={Stealth[length=2.2mm]},
    every node/.style={font=\small},
  ]
  \begin{scope}[scale=2.9]
    \draw[thick, black!70] (0,0) circle (1);
    \draw[black!70] (0.99,0) -- (1.04,0);
    \node[anchor=west] at (1.06,0) {$0$};
    \draw[->, black!55] (1.06,0.04) arc[start angle=1, end angle=14, radius=1.062];
    \fill[gwpurple, fill opacity=0.18] (-0.24,0) circle (0.3);
    \fill[gwcoral,  fill opacity=0.18] (0.24,0) circle (0.3);
    \begin{scope}
      \clip (-0.24,0) circle (0.3);
      \fill[gwteal, fill opacity=0.55] (0.24,0) circle (0.3);
    \end{scope}
    \draw[black!80, dashed] plot coordinates {(-0.540,-0.000) (-0.535,-0.056) (-0.517,-0.115) (-0.488,-0.169) (-0.452,-0.212) (-0.409,-0.248) (-0.355,-0.277) (-0.296,-0.295) (-0.240,-0.300) (-0.106,-0.288) (-0.006,-0.285) (0.095,-0.288) (0.249,-0.300) (0.301,-0.294) (0.346,-0.281) (0.385,-0.263) (0.424,-0.237) (0.459,-0.205) (0.485,-0.173) (0.507,-0.136) (0.525,-0.093) (0.536,-0.052) (0.540,-0.005) (0.536,0.052) (0.519,0.110) (0.493,0.161) (0.455,0.209) (0.409,0.248) (0.359,0.275) (0.305,0.293) (0.245,0.300) (0.106,0.288) (0.006,0.285) (-0.095,0.288) (-0.249,0.300) (-0.301,0.294) (-0.346,0.281) (-0.389,0.261) (-0.424,0.237) (-0.459,0.205) (-0.485,0.173) (-0.507,0.136) (-0.525,0.093) (-0.536,0.052) (-0.540,0.005)} -- cycle;
    \draw[gwpurple, thick] (-0.24,0) circle (0.3);
    \draw[gwcoral,  thick] (0.24,0) circle (0.3);
    \gwgeodesic[->, thick, gwblue!80!black]{148.00}{231.24}
    \node at (-0.33,0) {$K_1$};
    \node at (0.33,0) {$K_2$};
    \node[font=\scriptsize] at (0,0) {$K_0$};
    \node[anchor=north] at (0,-0.320) {$K_3$};
  \end{scope}

  \begin{scope}[shift={(5.6,-3)}]
    \fill[black!10] plot coordinates {(0.000,4.697) (0.100,4.766) (0.200,4.819) (0.300,4.856) (0.400,4.879) (0.517,4.891) (0.633,4.888) (0.750,4.874) (0.883,4.846) (1.033,4.802) (1.183,4.747) (1.350,4.676) (1.517,4.596) (1.900,4.392) (2.467,4.062) (2.483,4.080) (2.667,4.338) (2.800,4.504) (2.933,4.640) (3.000,4.697) (3.067,4.745) (3.133,4.785) (3.200,4.819) (3.267,4.845) (3.350,4.869) (3.433,4.884) (3.517,4.891) (3.600,4.890) (3.683,4.884) (3.833,4.858) (4.000,4.813) (4.167,4.754) (4.367,4.668) (4.583,4.563) (4.817,4.439) (5.100,4.278) (5.467,4.062) (5.483,4.080) (5.650,4.316) (5.767,4.465) (5.883,4.593) (6.000,4.697) (6.000,6.000) (0.000,6.000) (0.000,4.697)} -- cycle;
    \fill[gwblue, fill opacity=0.45] plot coordinates {(0.000,3.632) (0.167,3.592) (0.333,3.540) (0.517,3.472) (0.817,3.350) (1.000,3.289) (1.117,3.260) (1.217,3.242) (1.317,3.231) (1.400,3.229) (1.483,3.234) (1.567,3.246) (1.650,3.268) (1.733,3.299) (1.850,3.360) (2.000,3.465) (2.100,3.526) (2.200,3.575) (2.317,3.618) (2.433,3.647) (2.550,3.663) (2.700,3.667) (2.817,3.660) (3.017,3.629) (3.250,3.567) (3.467,3.492) (3.833,3.344) (4.017,3.285) (4.117,3.260) (4.217,3.242) (4.317,3.231) (4.400,3.229) (4.483,3.234) (4.567,3.246) (4.650,3.268) (4.717,3.292) (4.833,3.350) (5.017,3.476) (5.117,3.535) (5.233,3.589) (5.367,3.632) (5.500,3.657) (5.650,3.667) (5.817,3.660) (6.000,3.632) (6.000,3.750) (5.483,4.052) (5.467,4.055) (5.300,3.818) (5.183,3.667) (5.050,3.521) (4.933,3.418) (4.817,3.340) (4.750,3.306) (4.683,3.279) (4.550,3.243) (4.483,3.234) (4.400,3.229) (4.317,3.231) (4.233,3.239) (4.067,3.271) (3.883,3.326) (3.667,3.410) (3.450,3.510) (3.217,3.631) (2.917,3.798) (2.483,4.052) (2.467,4.055) (2.317,3.841) (2.217,3.708) (2.100,3.572) (2.000,3.474) (1.900,3.394) (1.783,3.322) (1.667,3.273) (1.550,3.243) (1.483,3.234) (1.400,3.229) (1.250,3.237) (1.083,3.267) (0.917,3.315) (0.717,3.389) (0.517,3.478) (0.283,3.595) (0.000,3.750) (0.000,3.632)} -- cycle;
    \draw[gwteal,   line width=1.2pt] plot coordinates {(0.000,3.632) (0.167,3.592) (0.333,3.540) (0.517,3.472) (0.800,3.356) (0.983,3.294) (1.100,3.264) (1.200,3.244) (1.300,3.232) (1.383,3.229) (1.467,3.232) (1.533,3.240) (1.617,3.258) (1.683,3.279) (1.750,3.306) (1.817,3.340) (2.050,3.497) (2.200,3.575) (2.283,3.608) (2.367,3.632) (2.450,3.650) (2.550,3.663) (2.683,3.667) (2.800,3.662) (2.933,3.645) (3.067,3.618) (3.217,3.578) (3.367,3.529) (3.533,3.466) (3.800,3.356) (3.983,3.294) (4.100,3.264) (4.200,3.244) (4.300,3.232) (4.383,3.229) (4.467,3.232) (4.550,3.243) (4.633,3.263) (4.717,3.292) (4.833,3.350) (5.017,3.476) (5.117,3.535) (5.233,3.589) (5.367,3.632) (5.500,3.657) (5.650,3.667) (5.817,3.660) (6.000,3.632)};
    \draw[gwpurple, line width=1.2pt] plot coordinates {(0.000,3.750) (0.283,3.595) (0.550,3.463) (0.783,3.363) (0.983,3.294) (1.083,3.267) (1.183,3.247) (1.350,3.230) (1.500,3.236) (1.583,3.250) (1.650,3.268) (1.767,3.314) (1.883,3.382) (2.000,3.474) (2.100,3.572) (2.183,3.667) (2.283,3.795) (2.600,4.247) (2.717,4.404) (2.850,4.559) (2.967,4.670) (3.033,4.722) (3.100,4.766) (3.167,4.803) (3.250,4.839) (3.317,4.860) (3.400,4.879) (3.483,4.889) (3.550,4.891) (3.633,4.888) (3.733,4.877) (3.833,4.858) (3.933,4.833) (4.150,4.760) (4.383,4.661) (4.617,4.546) (4.900,4.392) (6.000,3.750)};
    \draw[gwcoral,  line width=1.2pt] plot coordinates {(0.000,4.697) (0.100,4.766) (0.217,4.826) (0.333,4.865) (0.450,4.886) (0.583,4.891) (0.717,4.879) (0.867,4.850) (1.033,4.802) (1.183,4.747) (1.333,4.683) (1.500,4.605) (1.700,4.502) (2.050,4.307) (3.033,3.732) (3.300,3.587) (3.533,3.470) (3.783,3.363) (4.000,3.289) (4.100,3.264) (4.200,3.244) (4.300,3.232) (4.383,3.229) (4.467,3.232) (4.550,3.243) (4.633,3.263) (4.717,3.292) (4.783,3.322) (4.867,3.371) (4.933,3.418) (5.000,3.474) (5.117,3.590) (5.250,3.751) (5.367,3.911) (5.617,4.270) (5.750,4.445) (5.883,4.593) (6.000,4.697)};
    \draw[black, line width=1.0pt, dash pattern=on 4pt off 2.5pt] plot coordinates {(0.000,4.697) (0.100,4.766) (0.200,4.819) (0.300,4.856) (0.400,4.879) (0.517,4.891) (0.633,4.888) (0.750,4.874) (0.883,4.846) (1.033,4.802) (1.183,4.747) (1.350,4.676) (1.517,4.596) (1.900,4.392) (2.467,4.062) (2.483,4.080) (2.633,4.293) (2.733,4.425) (2.833,4.541) (2.933,4.640) (3.033,4.722) (3.150,4.794) (3.267,4.845) (3.383,4.876) (3.467,4.887) (3.550,4.891) (3.633,4.888) (3.733,4.877) (3.833,4.858) (3.933,4.833) (4.167,4.754) (4.417,4.645) (4.700,4.502) (5.033,4.316) (5.467,4.062) (5.483,4.080) (5.650,4.316) (5.767,4.465) (5.883,4.593) (6.000,4.697)};
    \draw[black!40, densely dotted] (0,3) -- (6,3);
    \draw[black!60, dashed] (0,0) -- (6,0) (0,6) -- (6,6);
    \draw[black!60] (0,0) -- (0,6);
    \draw[black!60, dotted] (6,0) -- (6,6);
    \fill[gwblue!80!black] (2.467,3.854) circle (0.07);
    \foreach \y/\lab in {0/0, 3/\pi, 6/2\pi}
      \node[anchor=east] at (-0.08,\y) {$\lab$};
    \foreach \x/\lab in {0/0, 3/\pi, 6/2\pi}
      \node[anchor=north] at (\x,-0.08) {$\lab$};
    \node at (3,-0.75) {$p$};
    \node[anchor=south] at (0,6.15) {$s$};
    \begin{scope}[shift={(0,-1.45)}]
      \draw[gwteal,   line width=1.2pt] (0,0) -- (0.5,0)   node[right, black] {$s_{K_0}$};
      \draw[gwpurple, line width=1.2pt] (1.6,0) -- (2.1,0) node[right, black] {$s_{K_1}$};
      \draw[gwcoral,  line width=1.2pt] (3.2,0) -- (3.7,0) node[right, black] {$s_{K_2}$};
      \draw[black, line width=1.0pt, dash pattern=on 4pt off 2.5pt] (4.8,0) -- (5.3,0) node[right, black] {$s_{K_3}$};
      \fill[black!10] (0,-0.62) rectangle (0.5,-0.38);
      \node[right] at (0.5,-0.5) {$W(K_3)$};
      \fill[gwblue, fill opacity=0.45] (3.2,-0.62) rectangle (3.7,-0.38);
      \node[right] at (3.7,-0.5) {$D_1 \cap D_2$};
    \end{scope}
  \end{scope}
\end{tikzpicture}
 \caption{Strong subadditivity for two overlapping connected wedges. Left: the
wedges $K_1$ and $K_2$, their meet $K_0 = K_1 \cap K_2$ and their join $K_3$,
the convex hull of $K_1 \cup K_2$ (dashed). The boundary point $p = 0$ is
marked, and $p$ increases counterclockwise. The geodesic shown cuts $K_1$ and
$K_2$ but misses $K_0$. Right: the critical angles $s_{K_i}(p)$. Each wedge
set $W(K_i)$ lies above its curve. $W(K_3) = W(K_1) \cap W(K_2)$ (shaded)
lies above $s_{K_3} = \max(s_{K_1}, s_{K_2})$. The configuration is symmetric
under rotation by $\pi$, so $s_{K_2}(p) = s_{K_1}(p + \pi)$. The difference set
$D_i$ is the band between $s_{K_0}$ and $s_{K_i}$. Their overlap
$D_1 \cap D_2$, whose measure is the strong subadditivity surplus in
Eq.~\eqref{eq:SSAresult} is shaded in blue. The blue
geodesic on the left is a sample geodesic whose wedge contains $K_0$ but not $K_1, K_2$. It corresponds to the point marked inside $D_1 \cap D_2$.}
  \label{fig:ssa}
\end{figure*}

\subsection{A difference formula}

To show monotonicity and strong subadditivity we first note the following result.  

\vspace{1mm}
\noindent \textbf{Lemma:}{ \it Let $K_1\subseteq K_2$ be connected convex wedges. Then,
\begin{equation}
  S_{\mathrm{gen}}(K_2)-S_{\mathrm{gen}}(K_1)
  =\nu\big(\mathcal{W}(K_1)\setminus\mathcal{W}(K_2)\big).
  \label{eq:master}
\end{equation}
}

To see this, note first that $\Gamma_{K_1}\subseteq\Gamma_{K_2}$, and
that a geodesic cutting $K_2$ but not $K_1$ leaves $K_1$ entirely on one
side. By \eqref{eq:trichotomy} the set of such geodesics is
\begin{equation}
  \Gamma_{K_2}\setminus\Gamma_{K_1}
  =\big(\mathcal{W}(K_1)\setminus\mathcal{W}(K_2)\big)
  \cup
  \big(\tilde{\mathcal{W}}(K_1)\setminus\tilde{\mathcal{W}}(K_2)\big).
\end{equation}
The above is 
a disjoint union of two sets which are exchanged under the transposition $(p,q)\to(q,p)$ and therefore are of equal measure. Taking the difference of \eqref{eq:SgenConvex} for the
two regions gives \eqref{eq:master}.

The set $(\mathcal{W}(K_1)\setminus\mathcal{W}(K_2))$ consists of the boundary
intervals whose entanglement wedge contains $K_1$ but not $K_2$. More explicitly, the set can be characterized on the kinematic space, following
\eqref{eq:fibre} as the band
\begin{equation}
  \mathcal{W}(K_1)\setminus\mathcal{W}(K_2)
  =\bigl\{(p,q) : s_{K_1}(p)<q-p<s_{K_2}(p)\bigr\}.
  \label{eq:band}
\end{equation}
between the two tangent curves. Since $K_1$ and $K_2$ are bounded away
from the conformal boundary, the two curves are bounded away from $s=0$
and $s=2\pi$, where $h$ diverges. The measure of the band \eqref{eq:band} is therefore
finite.
This is the kinematic space counterpart of the familiar statement that
differences of generalized entropies are finite while the entropies
themselves may not be, as in the case of standard entanglement wedges.

\subsection{Monotonicity}

For connected wedges monotonicity follows immediately from \cref{eq:master}. If $K_1\subseteq K_2$
then $\mathcal{W}(K_2)\subseteq\mathcal{W}(K_1)$, while the right hand side of
\eqref{eq:master} is the measure of a set. Positivity of $\nu$ thus gives
\begin{equation}
  S_{\mathrm{gen}}(K_2)\geq S_{\mathrm{gen}}(K_1).
\end{equation}
For wedges with several components, let $W_1=\bigsqcup_i A_i$ and
$W_2=\bigsqcup_j B_j$ with $W_1\subseteq W_2$. For each $j$ let
$S_j=\{i : A_i\subseteq B_j\}$. Each component of $W_1$ lies in exactly one
component of $W_2$, so the sets $S_j$ partition the components of $W_1$.

In \cref{sec:wedge_lattice}, we represented the lattice structure of generalized wedges on the kinematic space and saw that inclusion in the bulk reverses inclusion of corresponding wedge sets in kinematic space. Hence
$\mathcal{W}(B_j)\subseteq\mathcal{W}(A_i)$ for every $i\in S_j$, and therefore
\begin{equation}
  \mathcal{W}(B_j)\subseteq\bigcap_{i\in S_j}\mathcal{W}(A_i)=\mathcal{W}(H_j),
\end{equation}
where $H_j$ is the convex hull of $\{A_i\}_{i\in S_j}$. The equality is a consequence of the join property in the kinematic space.  The difference formula \eqref{eq:master} then gives
\begin{equation}
  S_{\mathrm{gen}}(B_j)-S_{\mathrm{gen}}(H_j)
  =\nu\big(\mathcal{W}(H_j)\setminus\mathcal{W}(B_j)\big)\geq 0 .
\end{equation}
Since $W_1$ is a generalized wedge, the characterization theorem requires the
hull of any collection of its components to have at least their total
generalized entropy,
\begin{equation}
  S_{\mathrm{gen}}(H_j)\geq\sum_{i\in S_j}S_{\mathrm{gen}}(A_i).
\end{equation}
Combining the two inequalities and summing over $j$ with
\eqref{eq:SgenGeneral} gives $S_{\mathrm{gen}}(W_1)\leq S_{\mathrm{gen}}(W_2)$.

Notice that the proof hinges on the difference formula \cref{eq:master} which in turn relies on the positivity of the measure $\nu$. By \eqref{eq:cmi}, positivity of $\nu$ follows directly from strong subadditivity of boundary intervals; in fact the surplus is an integral of conditional mutual
information of infinitesimally adjacent intervals \cref{eq:cmi} over the band \eqref{eq:band}. Monotonicity of the
generalized entropy is thus a consequence of strong subadditivity of the boundary
entropy.

\subsection{Strong subadditivity}

Let $K_1$ and $K_2$ be connected convex wedges, and write
$K_3=K_1\vee K_2$ and $K_0=K_1\wedge K_2$. All four wedge sets are
nested, with $\mathcal{W}(K_3)\subseteq\mathcal{W}(K_{1,2})\subseteq
\mathcal{W}(K_0)$. Define the difference sets
\begin{equation}
  D_i=\mathcal{W}(K_0)\setminus\mathcal{W}(K_i),
  \qquad i=1,2,3,
\end{equation}
each of finite measure by the argument following \eqref{eq:band}, and
each equal by \eqref{eq:master} to the corresponding entropy difference,
\begin{equation}
  \nu(D_i)=S_{\mathrm{gen}}(K_i)-S_{\mathrm{gen}}(K_0).
  \label{eq:Di}
\end{equation}
The join property of Sec.~\ref{sec:wedge_lattice} gives
$\mathcal{W}(K_3)=\mathcal{W}(K_1)\cap\mathcal{W}(K_2)$, and hence
\begin{equation}
  D_3=\mathcal{W}(K_0)\setminus
  \big(\mathcal{W}(K_1)\cap\mathcal{W}(K_2)\big)
  =D_1\cup D_2 .
  \label{eq:D3}
\end{equation}
The measure of a union is the sum of the measures of its parts, minus that of
their overlap, which would otherwise be counted twice. Thus
\begin{equation}
  \nu(D_3)=\nu(D_1)+\nu(D_2)-\nu(D_1\cap D_2).
\end{equation}
Substituting the entropy differences from \cref{eq:Di} we obtain
\begin{equation}
  S_{\mathrm{gen}}(K_1)+S_{\mathrm{gen}}(K_2)
  -S_{\mathrm{gen}}(K_3)-S_{\mathrm{gen}}(K_0)
  =\nu(D_1\cap D_2),
  \label{eq:SSAresult}
\end{equation}
where the right hand side is the measure of a set and therefore
non-negative. Thus we have shown strong subadditivity for connected GWs $K_1, K_2$.
For an illustration of the proof see \cref{fig:ssa}.

The details of the proof for GWs with several connected components require careful book-keeping for the intersections between various components, and a detailed proof is given
in Appendix \ref{app:SSA_disconnected}.

The identity \eqref{eq:SSAresult} also determines when strong subadditivity
is saturated. The set $D_1\cap D_2$ consists of boundary intervals whose
entanglement wedge contains $K_0$ but neither $K_1$ nor $K_2$. Since $K_0$ is
nonempty and each $K_i$ is connected, the RT geodesic of such an interval cuts
both $K_1$ and $K_2$ while missing $K_0$. In the vacuum the density
$h(p,q)$ is strictly positive, so $\nu(D_1\cap D_2)$ vanishes only when no
open family of such geodesics exists. This happens precisely when
$K_1\cup K_2$ is convex. If the union is convex, a geodesic meeting $K_1$ at
$x$ and $K_2$ at $y$ contains the segment $xy$, which lies in the union. The
intersections of the segment with $K_1$ and $K_2$ are overlapping intervals,
and the geodesic meets $K_0$. If the union is not convex, some point $z$ of
$\mathrm{hull}(K_1\cup K_2)$ lies outside the closure of $K_1\cup K_2$, on a
segment $xy$ with $x\in K_1$ and $y\in K_2$. The geodesic through $x$ and $y$
meets $K_1$ and $K_2$ in intervals separated by $z$, and so misses $K_0$.
Small displacements of $x$ and $y$ preserve this gap and give an open set of
such geodesics. Strong subadditivity is therefore saturated exactly when the
join coincides with the union, $K_1\vee K_2=K_1\cup K_2$.

Two remarks are in order. First, if $K_0$ is a proper subset of
$K_1\wedge K_2$ rather than equal to it, the inequality only becomes
stronger, by the monotonicity established above. Second, the argument
used no property of the vacuum state beyond \eqref{eq:hfromS} and
positivity, so it applies verbatim to any static holographic geometry
for which the Crofton density is a second derivative of a boundary
entropy.

\section{Discussion}
\label{sec:discussion}

We have shown that, in vacuum AdS$_3$ and at leading order in $G$, the lattice and entropic properties of generalized entanglement wedges admit a boundary description. The characterization of \cref{sec:characterization} expresses every generalized wedge through ordinary entanglement wedges. The space of ordinary entanglement wedges is the kinematic space, in which each GW is represented by a wedge set. Lattice operations on GWs are then represented as the corresponding lattice operations on wedge sets. Kinematic space also carries a measure, the Crofton form, and the generalized entropy of a GW is the measure of its wedge set. The key observation of this article is that the Crofton form is built from the von Neumann entropies of intervals in the dual CFT$_2$, and that its positivity follows from the strong subadditivity of these entropies. Monotonicity and strong subadditivity in the bulk follow from the positivity of the measure, and are therefore direct consequences of strong subadditivity in the dual CFT.

\paragraph*{Assumptions and limitations:} Our analysis keeps only the area term in the generalized entropy. This is the leading contribution as $G \to 0$ in the vacuum. At the next order, the generalized entropy of a wedge includes the entropy of bulk quantum fields, and the wedge itself is determined by a quantum extremal surface. The Crofton form is defined in terms of exact boundary entropies, so it already contains quantum corrections. Its integral over the geodesics crossing a closed curve need not, however, reproduce the bulk entropy of the region the curve encloses. The quantum corrections to the Ryu--Takayanagi formula concern entanglement across a surface anchored on the boundary \cite{Faulkner:2013ana}. The bulk entropy of a compact region involves entanglement across a closed surface that no single boundary interval computes. Whether a corrected form of differential entropy captures this contribution is an open question.
 
The assumption of vacuum AdS$_3$ is used in two places. First, the characterization theorem relies on the geometry of the hyperbolic plane. In particular, it uses the existence of a unique geodesic between any two points and the Crofton formula for convex sets. Second, the kinematic space construction relies on every bulk geodesic being a minimal surface anchored on the boundary. With matter or in quotients of AdS$_3$, such as conical defects and the BTZ black hole, some bulk geodesics are not minimal. Regions of the bulk may then lie in entanglement shadows that no Ryu--Takayanagi surface reaches. Their lengths are captured instead by nonminimal geodesics, and it has been proposed that their boundary interpretation involves entwinement rather than entanglement entropy \cite{Balasubramanian:2014sra}. Generalized wedges near such regions are therefore not directly described by our construction.

 

\paragraph*{Outlook:}Strong subadditivity is a property associated with subsystems of a quantum system. In our example, strong subadditivity of bulk GWs in AdS$_3$ follows directly from strong subadditivity of the dual CFT. This lends support to the idea that generalized entanglement wedges may also correspond to subsystems of the fundamental theory, which in this case is the CFT. Our results do not directly indicate what these subsystems are. In particular, we have not identified a density matrix of the CFT for the GWs. However, the structure of our proof provides some insights.

The generalized entropy of a GW corresponds to the \textit{differential entropy} of a family of intervals. This is not the von Neumann entropy of a boundary density matrix associated with a single subregion. Furthermore, the characterization result associates each GW with a wedge set in the kinematic space, which is equivalent to a collection of subregions of the boundary. These clues indicate that if there is a subsystem of the boundary CFT corresponding to a GW, it must be more general than a single boundary subregion.

This intuition supports the proposal of \cite{Sahu:2025upe}, which explores the possibility of assigning a subalgebra to every generalized wedge. The subsystems associated with GWs might thus be algebraic subsystems rather than subregions of the boundary CFT. In recent work \cite{Sahu:2026xyt, Sahu:2026rwk} we tested this proposal in tensor network models, and investigated when the family of generalized wedges can be consistently assigned a net of algebras.

If such an assignment exists, the characterization of \cref{sec:characterization} identifies two candidates for the algebra of a connected GW. Firstly, a connected GW is the intersection of all the entanglement wedges containing it. Secondly, it is also the intersection of the complements of all the entanglement wedges that do not intersect it.
The first expression suggests that the algebra of a connected GW is the intersection of the algebras associated with all the entanglement wedges containing it. A GW with several connected components would then be assigned the algebra generated by the algebras of its components. Thus, in the AdS$_3$ case, we can characterize the algebra associated with any generalized entanglement wedge through intersections and joins of algebras associated with ordinary entanglement wedges. The key question is then the precise way that the algebra associated with an ordinary entanglement wedge should be defined in the ``soft'' large $N$ limit, where we still include gravitational effects.

The second expression leads to a different characterization. Assuming that algebras of disjoint wedges commute, the algebra associated with a connected GW must be contained in the intersection of the commutants of the algebras associated with all the entanglement wedges that do not intersect it. Equivalently, it is contained in the commutant of the union (which is the commutant of the join) of the algebras associated with each of these wedges.
A straightforward possibility is that the algebra of the connected GW is equal to this commutant. This is closely related to the discussion of Leutheusser and Liu in \cite{Leutheusser:2022bgi}, where it is proposed that certain bounded bulk subregions are associated with the commutant of ``time band'' algebras. These are algebras of operators associated with the union of the domains of dependence of all the ordinary entanglement wedges that do not intersect the bulk region.
A similar commutant construction for regional algebras has been studied in global de Sitter space \cite{ElSayed:2026kml}.
Whichever candidate is correct, its entropy should reproduce the generalized entropy of the GW. In \cite{Sahu:2026xyt}, two different algebraic RT formulae have been proposed. It would thus be interesting to see whether these algebraic formulae reduce to the differential entropy.

Finally, a principal motivation for generalized wedges is that they do not require a conformal boundary. Our analysis uses the boundary in an essential way, since every quantity we compute is built from boundary entanglement entropies. It thus tests the Bousso--Penington proposal in a regime where an independent fundamental description exists. The agreement we find supports the view that the entropic properties of generalized wedges reflect genuine quantum information theoretic structure rather than a feature of the semiclassical minimization alone. Establishing an analogous statement in spacetimes without an asymptotic boundary, such as cosmologies, remains the main open problem.

\begin{acknowledgments}
I thank Mark Van Raamsdonk and Jeremy van der Heijden for useful discussions and comments on a draft of this paper.  Claude Opus 5.5 was used to proofread the manuscript, to check the technical statements and proofs in drafts written by the author, and to prepare the figures in TikZ. It also suggested ideas for the proof in Appendix A. 
\end{acknowledgments}

\appendix

\section{Strong subadditivity for disconnected wedges}
\label{app:SSA_disconnected}
Now we generalize the proof for generalized wedges with multiple connected components.
Let $K_1=\bigsqcup_i A_i$ and $K_2=\bigsqcup_m B_m$ be GWs with convex
components. Their meet is
\begin{equation}
  K_0=K_1\wedge K_2=\bigsqcup_{(i,m)\in E}\big(A_i\cap B_m\big),
\end{equation}
where $E$ is the set of overlapping pairs. Each $A_i$ and each $B_m$ lies in
exactly one connected component $K_3^{(a)}$ of the join $K_3=K_1\vee K_2$, and
overlapping pairs lie in the same component. 
Consider the union of the components of $K_1$ and $K_2$ inside a given $K_3^{(a)}$; the connected components of this union can be  organized into pieces $P$, and every pair in $E$ lies within a single piece.

The members of a piece $P$ can be ordered as $X_1,\dots,X_N$ so that each
$X_{n+1}$ overlaps an earlier member, since $P$ is connected. Let
$G_1=X_1$ and $G_{n+1}=\mathrm{hull}(G_n\cup X_{n+1})$. The wedges $G_n$ and
$X_{n+1}$ are connected, convex and overlapping, with join $G_{n+1}$ and meet
$G_n\cap X_{n+1}$. The connected result \eqref{eq:SSAresult} then gives
\begin{equation}
  S_{\mathrm{gen}}(G_n)+S_{\mathrm{gen}}(X_{n+1})
  \geq S_{\mathrm{gen}}(G_{n+1})+S_{\mathrm{gen}}(G_n\cap X_{n+1}).
\end{equation}
Summing the above inequality over $n$, the terms in $G_n$ cancel pairwise. Since $G_N$ is the hull
of $P$, we have
\begin{equation}
  \sum_{n=1}^{N}S_{\mathrm{gen}}(X_n)
  \geq S_{\mathrm{gen}}\big(\mathrm{hull}(P)\big)
  +\sum_{n=1}^{N-1}S_{\mathrm{gen}}(G_n\cap X_{n+1}).
  \label{eq:pieceSSA}
\end{equation}

The intersections on the right hand side contain the components of the meet.
If $X_{n+1}=B_m$, the convex set $G_n\cap B_m$ contains $A_i\cap B_m$ for
every earlier $A_i$ overlapping $B_m$, and the same holds with the roles of
$A$ and $B$ exchanged. These sets are components of the GW $K_0$, and by the
characterization theorem any collection of them is itself a GW. Monotonicity
then gives
\begin{equation}
  S_{\mathrm{gen}}(G_n\cap B_m)\geq\sum_{i\,\mathrm{earlier}}
  S_{\mathrm{gen}}(A_i\cap B_m).
\end{equation}
Each pair in $E$ enters exactly once, at the step where its later member
joins the piece. Substituting into \eqref{eq:pieceSSA} we obtain
\begin{equation}
  \sum_{X\in P}S_{\mathrm{gen}}(X)
  \geq S_{\mathrm{gen}}\big(\mathrm{hull}(P)\big)
  +\sum_{(i,m)\in E\cap P}S_{\mathrm{gen}}(A_i\cap B_m).
  \label{eq:pieceMeet}
\end{equation}

The hulls of the pieces inside $K_3^{(a)}$ are related to $K_3^{(a)}$ itself
through the minimality of the join. The region obtained from $K_3$ by
exchanging $K_3^{(a)}$ for the union of these hulls still contains
$K_1\cup K_2$. As the GW of $K_1\cup K_2$, the join has the smallest
generalized entropy of all such regions. The boundary of a union lies within
the union of the boundaries, and hence
\begin{equation}
  \sum_{P\subseteq K_3^{(a)}}S_{\mathrm{gen}}\big(\mathrm{hull}(P)\big) \geq S_{\mathrm{gen}}\big(K_3^{(a)}\big).
  \label{eq:joinBound}
\end{equation}
Summing \eqref{eq:pieceMeet} over all pieces and bounding the hulls by
\eqref{eq:joinBound}, and further summing over the components of $K_3$ indexed by $a$, we obtain
\begin{equation}
  S_{\mathrm{gen}}(K_1)+S_{\mathrm{gen}}(K_2)
  \geq S_{\mathrm{gen}}(K_3)+S_{\mathrm{gen}}(K_0).
\end{equation}
Thus strong subadditivity holds for GWs with any number of connected
components.

\bibliographystyle{apsrev4-2}
\bibliography{references}

\end{document}